\documentclass[10pt]{article}

\usepackage{amsmath}
\usepackage{amssymb}

\usepackage{graphicx}

\usepackage{cite}

\usepackage{color} 

\usepackage[labelfont=bf,labelsep=period,justification=raggedright]{caption}

\makeatletter
\renewcommand{\@biblabel}[1]{\quad#1.}
\makeatother

\date{}

\begin{document}

\begin{flushleft}
{\Large
\textbf{A compact statistical model of the song syntax in Bengalese finch}
}
\\
Dezhe Z. Jin$^{\ast}$ and Alexay A. Kozhevnikov
\\
\bf Department of Physics, The Pennsylvania State University, 
University Park, PA 16802, U.S.A.
\\
$\ast$ E-mail: Corresponding djin@phys.psu.edu
\end{flushleft}

\section*{Abstract}

Songs of many songbird species consist of variable sequences of a finite number of syllables. 
A common approach for characterizing the syntax of these complex syllable sequences is to use transition probabilities between the syllables.
This is equivalent to the Markov model, in which each syllable is associated with one state, and the transition probabilities between the states do not depend on the state transition history.
Here we analyze the song syntax in a Bengalese finch. 
We show that the Markov model fails to capture the statistical properties of the syllable sequences. 
Instead, a state transition model that accurately describes the statistics of the syllable sequences
includes adaptation of the self-transition probabilities when states are repeatedly revisited, and allows associations of more than one state to the same syllable. 
Such a model does not  increase the model complexity significantly. 
Mathematically, the model is a partially observable Markov model with adaptation (POMMA). 
The success of the POMMA supports the branching chain network hypothesis of how syntax is controlled within the premotor song nucleus HVC, 
and suggests that adaptation and many-to-one mapping from neural substrates  to syllables are important features of the neural control of complex song syntax. 

\section*{Author Summary}

Complex action sequences in many animals are organized according to syntactical rules. 
A critical problem for understanding the neural basis of action sequences is how to derive the syntax that captures the statistics of the sequences. 
Here we solve this problem for the songs of a Bengalese finch, which consists of variable sequences of several stereotypical syllables. 
The Markov model is widely used for describing variable birdsongs, 
where  each syllable is associated with one state, and the transitions between the states are stochastic and depend only on the state pairs. 
However, such a model fails to describe the syntax of the Bengalese finch song.
We show that two modifications are needed. The first is adaptation. Syllable repetitions are common in the Bengalese finch songs. 
Allowing the probability of repeating a syllable to decrease with the number of repetitions leads to better fits to the observed repeat number distributions. The second is many-to-one mapping from the states to the syllables. 
A given syllable can be generated by more than one state. 
Even if the transitions between the states are Markovian, the syllable statistics are not Markovian due to the multiple representations of the same syllables.
With these modifications, 
the model successfully describes the statistics of the observed syllable sequences, including the repeat number distributions, the probability of observing a given  syllable at a given step from the start, and the N-gram distributions. 

\section*{Introduction}

Complex action sequences in animals and humans are often organized according to syntactical rules \cite{Lashley1951,colonnese1996ontogeny}.
Many examples are found in birdsong.  
Songbird species such as Bengalese finch \cite{Woolley1997, Honda1999, Okanoya2004}, sedge warbler \cite{Catchpole1976}, 
nightingale \cite{Todt1998}, and willow warbler \cite{Gil2000} consist of a finite number of stereotypical syllables (or notes) 
arranged in variable sequences. 
Quantitative analysis of the action syntax is critical for understanding the neural mechanisms of how complex sequences are generated \cite{Lashley1951,Woolley1997,Okanoya2004,sakata2006real,Jin2009}, and for comparative studies of learning and cultural transmissions of sequential behaviors \cite{slater1989bird}.

Pairwise transition probabilities between syllables are widely used to characterize variable birdsong sequences \cite{Woolley1997, Honda1999, Todt1998, Gil2000}.
This is equivalent to the Markov model, 
in which each state is associated with one syllable, and the transition probabilities between the states depend only on the state pairs. 
More sophisticated models use finite state automata, in which transitions between states emit chunks of fixed sequences of syllables, 
with a possibility that a syllable appears in different chunks \cite{Hosino2000,Okanoya2004,kakishita2007pattern}.
However, no statistical tests of these models have been performed, 
and their validity as quantitative descriptions of the birdsong syntax remains unclear. 

In this paper, we analyze the songs of a Bengalese finch.
We demonstrate that the Markov model fails to capture the statistical properties of the observed sequences, including the repeat number distributions of individual syllables, the distributions of the N-grams (sequences of length N) \cite{Jurafsky2000} and the probability of observing a given syllable at a given step from the start. 
We show that two modifications are needed. The first is adaptation. Syllable repetitions are common in the Bengalese finch songs. 
Allowing the repeat probabilities of syllables to decrease with the number of repetitions leads to a better fit of the repeat number distributions. 
The second is many-to-one mapping from the states to the syllables. A given syllable can be generated by more than one state. Even if the transitions between the states are Markovian, the syllable statistics are not Markovian due to the multiple representations of the same syllables. 
The resulting model, which we call a partially observable Markov model
with adaptation  (POMMA), 
has history-dependent transition probabilities between the states and 
many-to-one mappings from the states to the syllables.

The POMMA successfully describes the statistical properties of the observed syllable sequences. The adaptation and many-to-one mapping are further supported by comparing the acoustic features of the same syllable appearing in different places in the sequences. The results are consistent with the branching chain network model of generating variable birdsong syntax, in which syllable-encoding chain networks of projection neurons in the premotor song nucleus HVC are connected in a branching topology \cite{Jin2009,Chang2009}.

\section*{Results}

\begin{figure}[!ht]
\begin{center}
\includegraphics[scale=1]{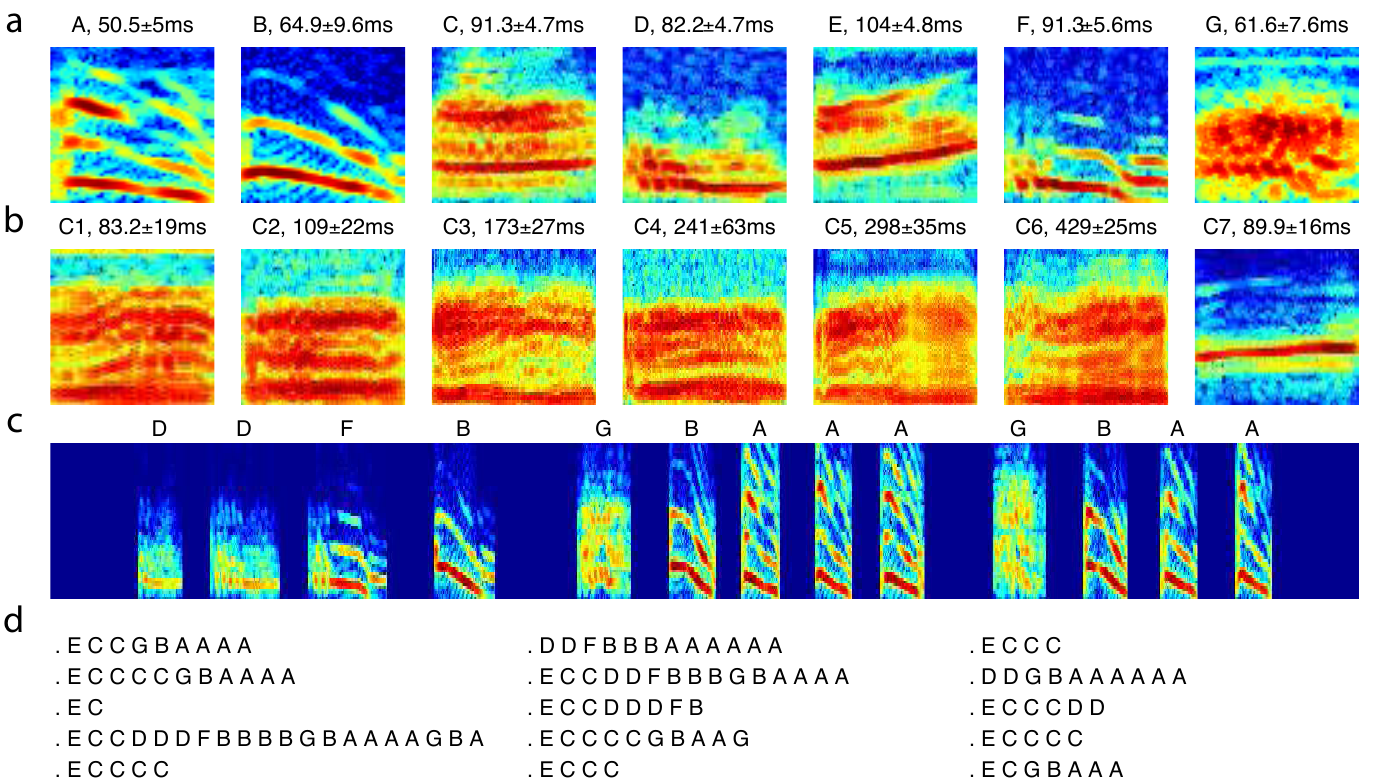}
\end{center}
\caption{
{\bf Spectrograms and song sequences.}
{\bf a}. Spectrograms of song syllable types. 
{\bf b}. Spectrograms of call types. 
{\bf c}. Spectrogram of an example song. 
Syllable types are shown on top. 
The duration of the song measured from the start of the first syllable to the end of the last syllable is 1.4s.
{\bf d}. Examples of the syllable sequences. 
The frequency range of the spectrograms are $1$-$10kHz$.
The durations of the syllable and call types are shown 
on top of the spectrograms.
}
\label{JinFigSyllablesAndSongs}
\end{figure}

Spontaneous vocalizations of a Bengalese finch were recorded in an acoustic chamber using a single microphone over six days. 
Vocal elements ($n=25365$) 
were isolated from the recorded pressure waves, 
and clustered into 25 types according to the similarities of 
their spectrograms (Materials and Methods). 
We identified seven types of vocal elements as song syllables (Fig.\ref{JinFigSyllablesAndSongs}a, $n=4625, 3145, 2835, 2154, 1408, 723, 1356$ for syllables A to G, respectively).
The rest were call notes (14 types; 7 examples shown in Fig.\ref{JinFigSyllablesAndSongs}b; C1 and C2 were the the most frequent call notes with $n=2200, 918$, respectively) and noise.  
The song syllables were distinguished by rich structures in the spectrograms 
and tight distributions of the durations (s.d./mean $=0.08\pm0.04$),  
(Fig.\ref{JinFigSyllablesAndSongs}a), 
and frequently appeared together in long sequences (sequence length mean 
$=8.5 \pm 4.9$) with small inter-syllable gaps ($<200 ms$)
(Fig.\ref{JinFigSyllablesAndSongs}d). 
In contrast, the call notes had broad or simple spectra and more variable distributions of the durations (s.d./mean $=0.17 \pm 0.05$), 
and appeared in short sequences (sequence length mean $=2.7 \pm 1.5$). 
All consecutive sequences of the song syllables with inter-syllable gaps smaller than $200ms$ were assigned as song sequences.
Additionally, syllable E (Fig.\ref{JinFigSyllablesAndSongs}a), which  predominantly appeared at the start of the sequences obtained above, 
was assigned as a start syllable such that whenever syllable E appeared for the first time and was not following another E, a new song sequence was started. 
Thus, a long sequence containing $k$ non-continuous E's in the middle was broken into $k+1$ song sequences. 
Altogether, we ended up with 1921 song sequences. 
Sequences of call notes can precede or follow song sequences, and these call notes were considered to be introductory notes. 

\subsection*{The Markov model}

\begin{figure}[!ht]
\begin{center}
\includegraphics[scale=1]{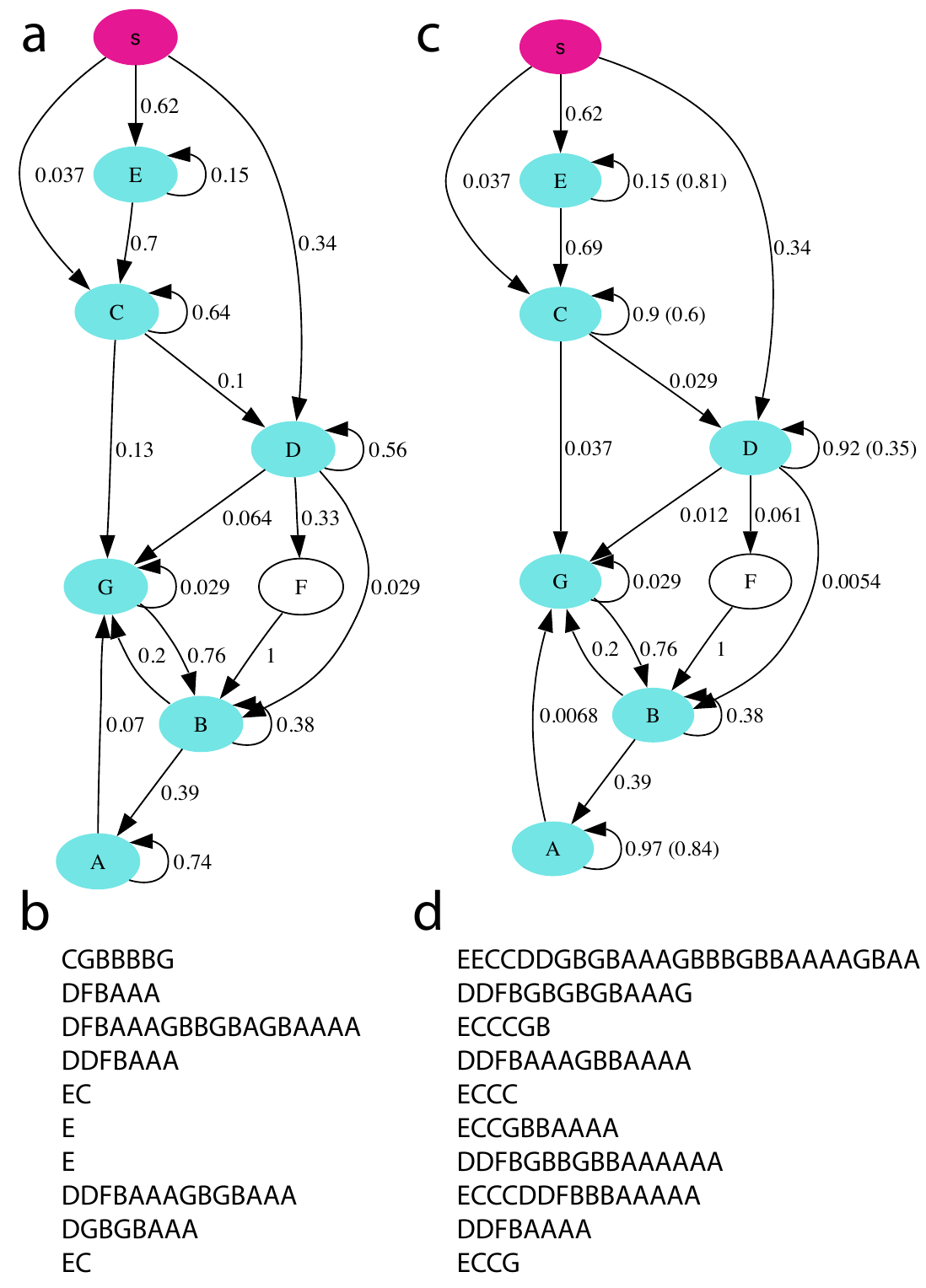}
\end{center}
\caption{
{\bf The Markov syntax of the syllable sequences.}
{\bf a}. The Markov model. 
The pink oval represents the start state.
Cyan indicates that the state has a finite probability of transitioning to the end state. The numbers indicates the transition probabilities. 
{\bf b}. Examples of syllable sequences generated from the Markov model. 
{\bf c}. The Markov model with adaptation. 
The numbers in parenthesis are the adaptation parameter $\alpha$.
{\bf d}. Examples of syllable sequences generated from the Markov model with adaptation. 
}
\label{JinFigSyntaxMarkov}
\end{figure}

A simple statistical model of the song sequences is the Markov model, which is completely specified by the transition probabilities between the syllables. 
For each syllable, there is a corresponding state; additionally, there is a start state (symbol $s$) and an end state 
(symbol $e$), as shown in Fig.\ref{JinFigSyntaxMarkov}a. 
We computed the transition probability $p_{ij}$ for the state $S_i$ associated with syllable $i$ to the state $S_j$ associated with syllable $j$, 
from the observed song sequences 
as the ratio of the frequency of the sequence $ij$ 
over the total frequency of syllable $i$.  
Transitions with small probabilities ($p_{ij} < 0.01$) were excluded. 

To evaluate how well the Markov model describes the statistics of the observed song sequences, 
we generated 10000 sequences from the model, and compared three statistical properties of the generated sequences and the observed sequences. 
The method of sequence generation is as follows. 
From the start state, one of three states $S_C, S_E, S_D$ associated with syllables C, E, D can be followed with probabilities $p_{sC}=0.037$, $p_{sE}=0.625$, $p_{sD}=0.338$, respectively (Fig.\ref{JinFigSyntaxMarkov}a). 
A random number $r$ is uniformly sampled from 0 to 1. 
If $r < p_{sC}$, $S_C$ is selected (the state following the start state is $S_1=S_C$), and the generated sequence starts with C. 
If $p_{sC} < r< p_{sC} + p_{sE}$, $S_E$ is selected ($S_1=S_E$), and the sequence starts with E. 
If $p_{sC} + p_{sE} < r < p_{sC} + p_{sE} + p_{sD} = 1 $, $S_D$ is selected ($S_1=S_D$), and the sequence starts with D. 
From the selected state $S_1$, the next state $S_2$ can be selected similarly according to the transition probabilities from $S_1$.
This process of sampling random numbers and selecting the next state and syllable is continued until the end state is reached, generating a specific syllable sequence.  
Examples of the generated syllable sequences are shown in Fig.\ref{JinFigSyntaxMarkov}b. 

\begin{figure}[!ht]
\begin{center}
\includegraphics[scale=1]{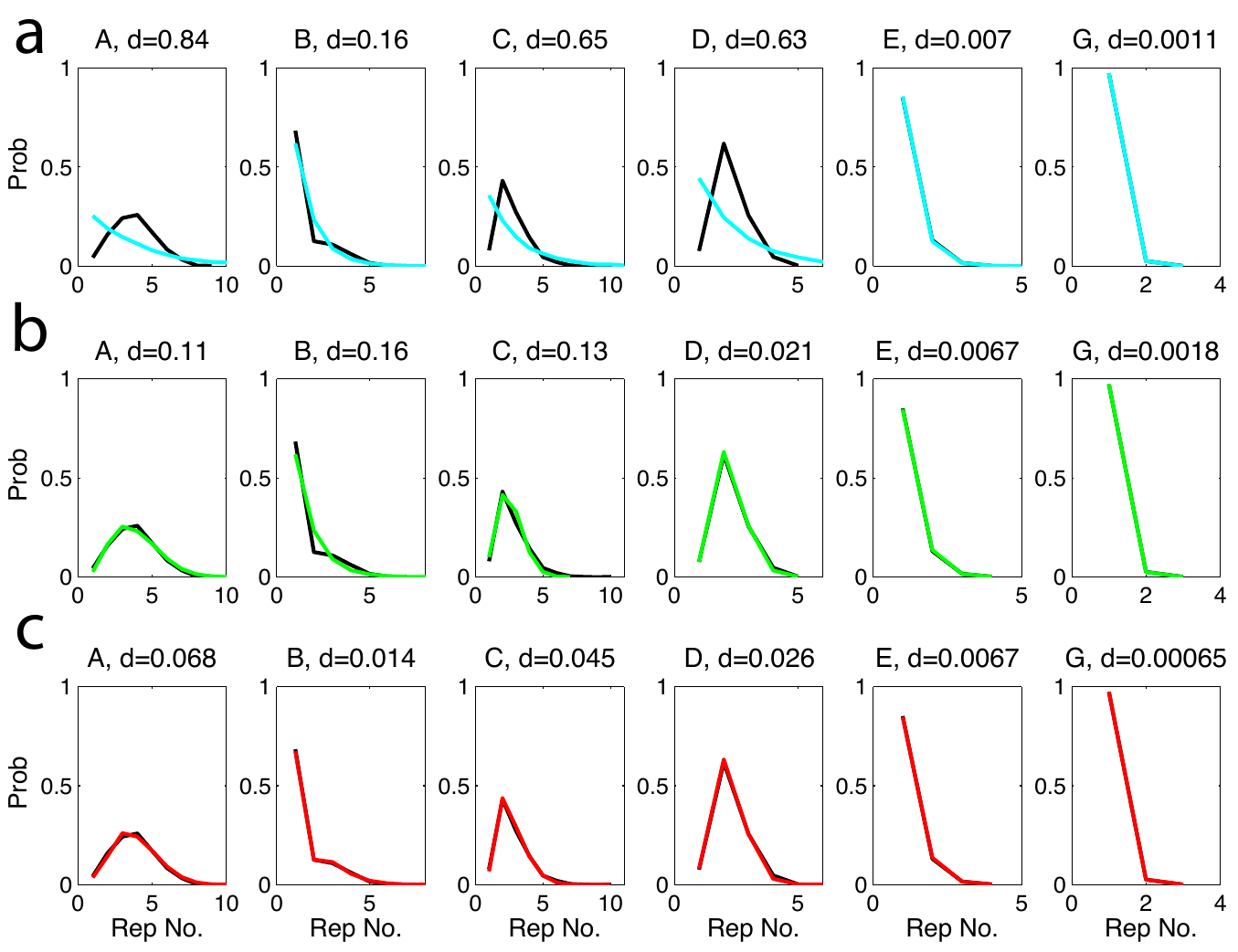}
\end{center}
\caption{
{\bf 
Comparisons of the repeat distributions for syllables A,B,C,D,E,G.
}
The black curve in each graph is from the observed syllable sequences. 
{\bf a}. Comparison to the distributions from the Markov model (cyan curves).
{\bf b}. Comparison to the distributions from the Markov model with 
adaptation (green curves).
{\bf c}. Comparison to the distributions from the POMMA (red curves). 
The differences between the model and the observed curves are indicated above each graph with the $d$-values.
}
\label{JinFigRepeatStats}
\end{figure}

The first statistical property compared was the distributions of the syllable repeats. 
Except syllable F, all syllables appeared in repetitions, and the number of repeats were variable. 
For each syllable, we constructed the probability distribution of the repeat numbers by counting the frequencies of observing a given number of repeats in the observed song sequences. 
The distributions are shown as black curves in Fig.\ref{JinFigRepeatStats}a. 
We also constructed the repeat number distributions from the sequences generated from the Markov model.
These are shown as cyan curves in Fig.\ref{JinFigRepeatStats}a.
For syllables E and G, the comparisons are favorable.
However, for other syllables the distributions clearly disagree. 
To quantify the difference between two distributions $f_1 (x)$ and $f_2(x)$, 
we defined the maximum normalized difference $d$, which is the 
maximum of the absolute differences divided by the maximum values in the two distributions, i.e. 
 $d = \max_x | f_1(x) - f_2(x) | / \max(f_1(x),f_2(x))$. 
 The $d$-values for syllables A, B, C, D, E are 
 $0.84, 0.16, 0.65, 0.63, 0.007, 0.0011$, respectively. 
The major difference is that, for syllables A, C, D, 
the observed distributions peak at repeat numbers 4, 2, 2, respectively, 
while the generated distributions are decreasing functions of the repeat numbers. 
Indeed, if the probability of returning to state $S$ from itself is $p$, 
the probability of observing $n$ repeats of the associated syllable is 
$P_n = p^{n-1} (1-p)$, which is a decreasing function of $n$. 
Therefore the Markov model is incapable of producing repeat number distributions
having maxima at $n > 1$. 

\begin{figure}[!ht]
\begin{center}
\includegraphics[scale=1]{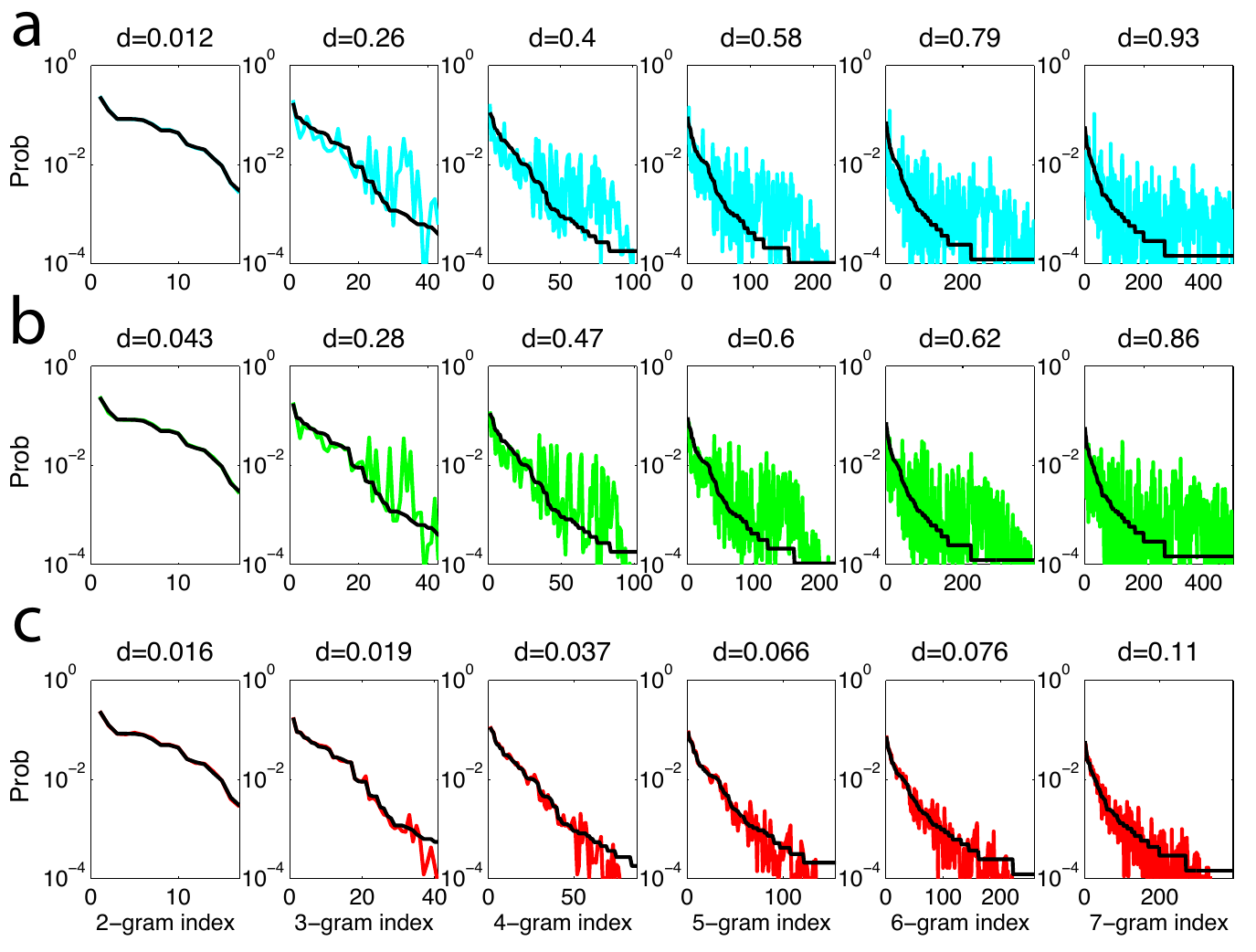}
\end{center}
\caption{
{\bf
Comparisons of the N-gram distributions.} 
{\bf a}. The Markov model. 
{\bf b}. The Markov model with adaptation. 
{\bf c}. The POMMA. 
The conventions are the same as in Fig.\ref{JinFigRepeatStats}. }
\label{JinFigNGramStats}
\end{figure}

The second statistical property compared was the N-gram distributions. 
An N-gram is a fixed subsequence of length $N$. 
For example, syllable sequences EC and AA are 2-grams; ECC and AAA are 3-grams; etc. 
We constructed the probability distributions for 2- to 7-grams in the observed song sequences by counting the frequencies of a given subsequence. 
The results are shown in Fig.\ref{JinFigNGramStats}a as black curves, 
with the N-grams sorted according to decreasing probabilities.  
We also computed the probability distributions of the corresponding N-grams in the generated sequences. 
The results are shown in Fig.\ref{JinFigNGramStats}a as cyan curves. 
The distributions for 2-grams agree very well, which is expected, 
since the Markov model was constructed with the transition probabilities, which are equivalent to the 2-gram distributions.  
The distributions are quite different for 3- to 7-grams, with $d$-values ranging from 0.26 to 0.93 (Fig.\ref{JinFigNGramStats}a). 

\begin{figure}[!ht]
\begin{center}
\includegraphics[scale=1]{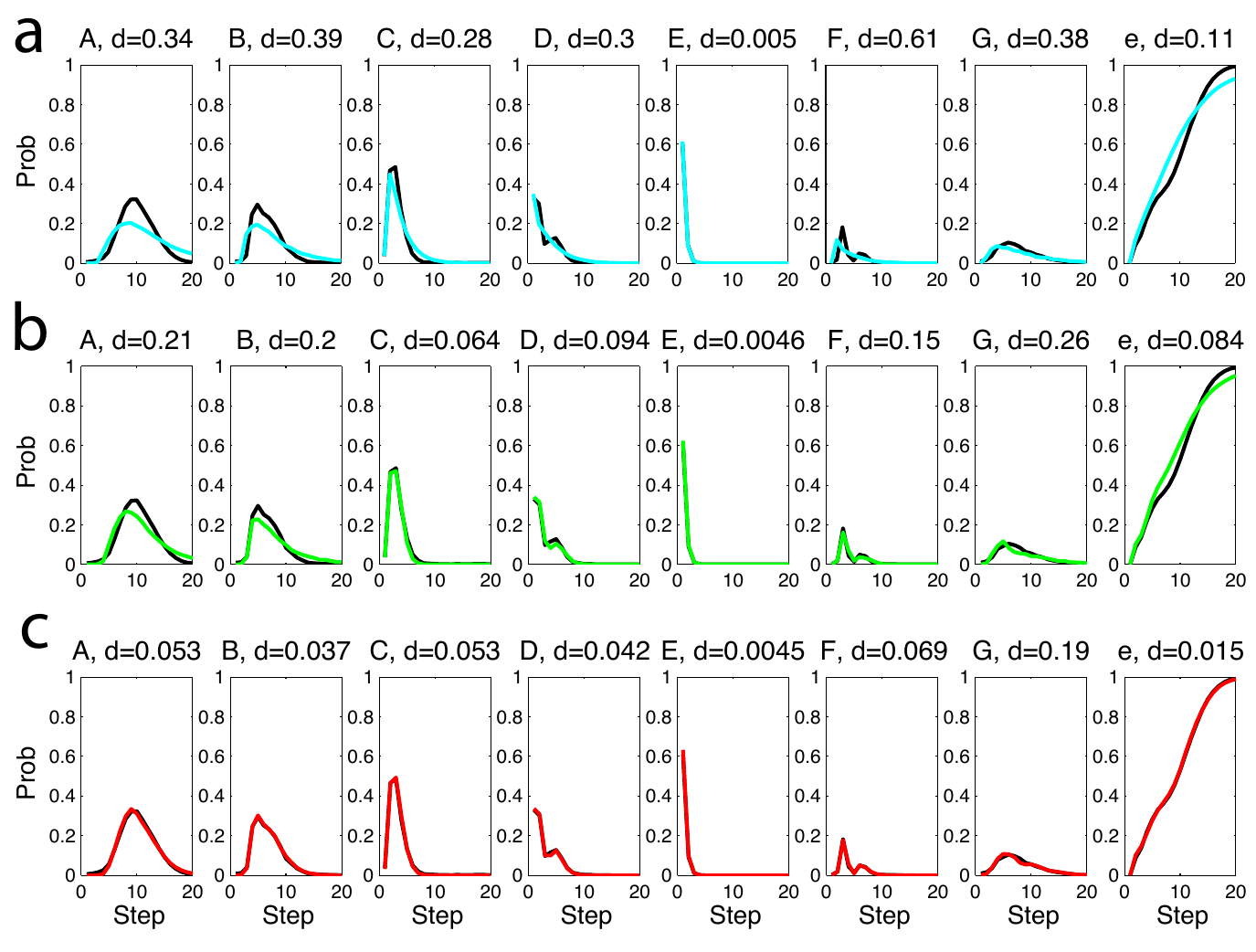}
\end{center}
\caption{
{\bf
Comparisons of the probabilities of finding the syllables and the end (denoted with e) at a given step from the start. }
{\bf a}. The Markov model.
{\bf b}. The Markov model with adaptation.
{\bf c}. The POMMA.
The conventions are the same as in Fig.\ref{JinFigRepeatStats}. 
}
\label{JinFigPNStepsStats}
\end{figure}

The final statistical property compared was the step probability of a syllable, defined as the probability of observing the syllable at a given step from the start. 
The step probabilities for all syllables
computed from the observed song sequences,
as well as the step probability of the end symbol $e$, which 
describes the probability of observing that a sequence has ended at or before a given step, or equivalently, the cumulative distribution function of the sequence length, 
are plotted as black curves in Fig.\ref{JinFigPNStepsStats}a; 
and those from the  generated sequences are plotted as cyan curves. 
The comparison for syllable E is quite good ($d=0.005$).
But the differences between the probabilities for other syllables and the end symbol $e$ are large, as indicated by 
the $d$-values ranging from 0.11 to  0.61. 

\begin{figure}[!ht]
\begin{center}
\includegraphics[scale=1]{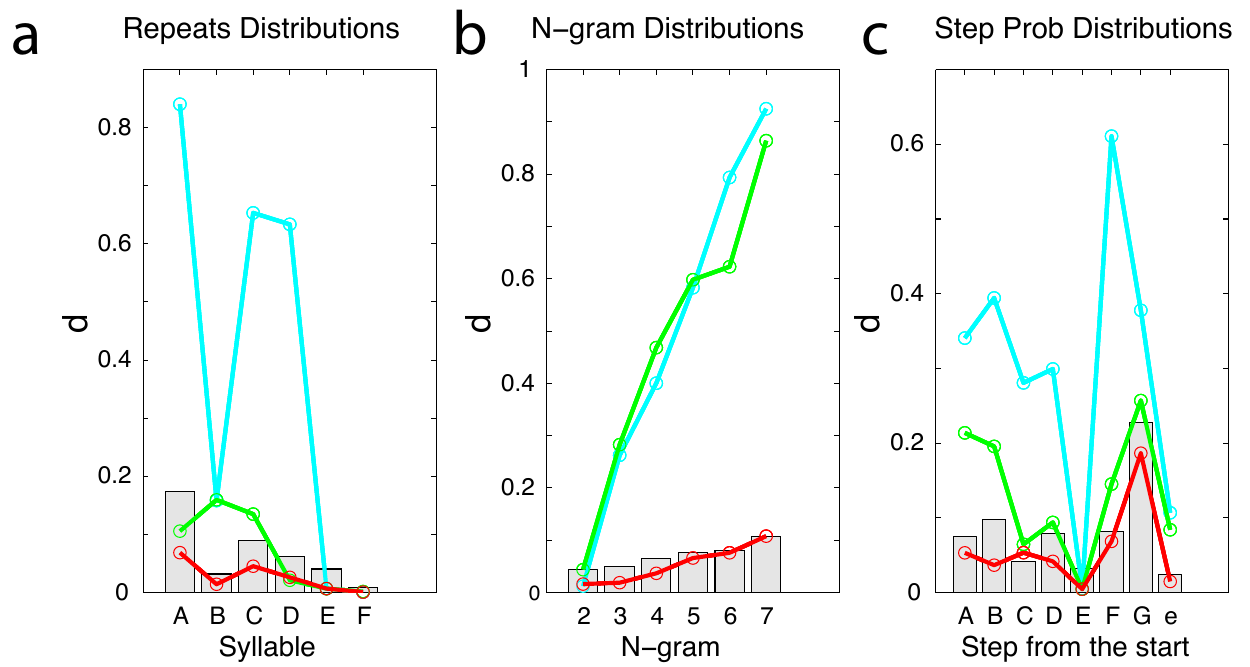}
\end{center}
\caption{
{\bf
Summary of the differences between the model-generated and observed distributions.
} 
Cyan lines are from the Markov model, green lines from the Markov model with adaptation, and the red lines from the POMMA. 
The gray bars are the estimates of the noisy fluctuations in the observed distributions. 
{\bf a}. The repeat distributions. 
{\bf b}. The N-gram distributions. 
{\bf c}. The probabilities of observing syllables and the end in a given step from the song start. 
}
\label{JinFigErrorsSummary}
\end{figure}

Because the number of the observed song sequences is finite, even a perfect statistical model of the observed song sequences cannot lead to perfect fits to the observed distributions.
To obtain reasonable benchmarks of the $d$-values expected from this finite size effect, we randomly assigned the observed song sequences into two groups, and computed the $d$-values by comparing the two groups. 
The random splitting was done 500 times, and the $d$-value distributions were constructed for the repeat number distributions, the N-gram distributions, and the step probabilities.  
The $p=0.95$ points in these distributions were selected as the benchmarks (i.e. 95\% of the $d$-values were smaller than the benchmarks).
They are plotted as gray vertical bars in Fig.\ref{JinFigErrorsSummary}. 
The $d$-values obtained from the Markov model, plotted as the cyan curves in Fig.\ref{JinFigErrorsSummary}, 
are mostly far beyond the benchmarks.  
It is clear that the Markov model fails to capture the statistical properties of the Bengalese finch song. 

\subsection*{The Markov model with adaptation} 

Adaptations are widely observed in neural systems. 
Continuous firing can reduce neuron excitability \cite{Sanchez2000}, 
and excitatory synapses can be less effective when activated repeatedly \cite{Markram1996,Abbott1997}. 
In zebra finch, consecutive singing lengthens the motif durations \cite{Chi2001}.
Recent experiments and computational work suggest that 
singing in zebra finch is driven by synaptic chain networks of HVC projection neurons \cite{Hahnloser2002,Jin2007,Long2008,Long2010};
it is possible that the motif slow-down is due to some adaptive processes in HVC.  
In the branched chain network model of the Bengalese song syntax, 
reduction in synaptic efficacy in HVC projection neurons reduces transition probability between the syllables \cite{Jin2009}. 
These observations suggest that the transition probabilities might not be fixed. 
We therefore tested whether introducing adaptation to the Markov model leads to a better statistical model for the observed song sequences. 

Ideally, all transition probabilities should be subject to dynamical changes depending on the histories of the state transitions in the Markov model. 
But such a model is difficult to analyze. 
We therefore considered a simple model in which 
only the return probabilities of the states from themselves are adaptive. 
In particular, the return probability $p_r$ of a state
is reduced to $p_r = \alpha^n p$ after $n$-th repetition of the associated syllable. 
The transition probabilities to all other states are mutipled by a factor 
$(1 - \alpha^n p)/(1-p)$ to keep the total probability normalized. 
Here $0<\alpha<1$ is the adaptation parameter, and $p$ is the return probability when $n=1$.   
The probabilities recover to original values once the dynamics moves onto other states. 
In this Markov model with adaptation, 
the probability of observing $n$ repetitions is given by 
$P_n = \alpha^{(n-2)(n-1)/2} p^{n-1} ( 1 - \alpha^{n-1} p) $
(Materials and Methods). 
We fitted the parameters $\alpha$ and $p$ for the states with self-transitions in the Markov model (Fig.\ref{JinFigSyntaxMarkov}a) using the repeat number distributions in the observed song sequences. 
The resulting model is shown in Fig.\ref{JinFigSyntaxMarkov}b, 
which is identical to the Markov model (Fig.\ref{JinFigSyntaxMarkov}a) except that the return probabilities for the states associated with syllables A, C, D, E are adaptive, with $\alpha=0.84, 0.6, 0.35, 0.81$, respectively.  
Fittings for syllables B and G did not lead to an adaptive model ($\alpha=1$), so the associated return probabilities are unchanged. 

To evaluate the Markov model with adaptation, we again generated 10000 song sequences and compared the repeat number distributions, the N-gram distributions, and the step probabilities to the observed song sequences. 
The generation procedure was the same as in the original Markov model, except that the return probabilities were adaptive as prescribed above. 
The repeat number distributions, shown as green curves in Fig.\ref{JinFigRepeatStats}b, are much improved compared to the Markov model. 
In particular, the peaked distributions of syllables A, C, D are well reproduced.   
This demonstrates that the adaptation is capable of producing peaked repeat number distributions.
Adaptation did not improve the comparisons of the N-gram distributions (Fig.\ref{JinFigNGramStats}b), 
and improved the comparisons of the step probabilities for syllables C, D, F but not for syllables A, B, D and the end symbol $e$ 
(Fig.\ref{JinFigPNStepsStats}b). 
The $d$-values (green curves in Fig.\ref{JinFigErrorsSummary}) compared to the benchmarks confirm these observations. 
The Markov model with adaptation is a better statistical model for the Bengalese song sequences than the Markov model; 
however, it is still not capable of accurately describing all statistical properties. 

\subsection*{Partially observable Markov model with adaptation (POMMA)}

In the Markov model and its extension with adaptation, each syllable is associated with one state. 
Hence the number of states equals to the number of the syllables, plus two if we count the start and end states. 
However, nothing excludes the possibility  
that there is more than one state corresponding to one syllable. 
This many-to-one mapping from the states to the syllables enables more elaborate statistical properties of the song sequences \cite{Jin2009}. 
Biologically, since HVC neurons drive singing by projecting
to RA (the robust nucleus of the arcopallium) \cite{Nottebohm1976,Nottebohm1982}, 
and these connections are learned \cite{Doya1995,Fee2004,Fiete2004}, 
different sets of HVC projection neurons can drive acoustically similar syllables. 
In zebra finch, different neural activities in HVC have been observed during vocalizations of acoustically similar syllables \cite{McCasland1987,Yu1996}, 
supporting the possibility that the same syllable types are represented with distinctive neural substrates. 
With the many-to-one mapping, the number of states can be larger than the number of syllables plus two.
When this is the case, 
some of the states are ``hidden", and cannot be simply deduced by counting the number of syllable types. 
This kind of model is often referred to as ``partially observable Markov model" (POMM) \cite{Callut2004}, and is a special case of hidden Markov model in which each state is associated with a single symbol. 
We tested whether introducing many-to-one mapping in addition to the adaptation, which leads to a ``partially observable Markov model with adaptation" (POMMA), would better explain the statistical properties of the observed song sequences.  

\begin{figure}[!ht]
\begin{center}
\includegraphics[scale=1]{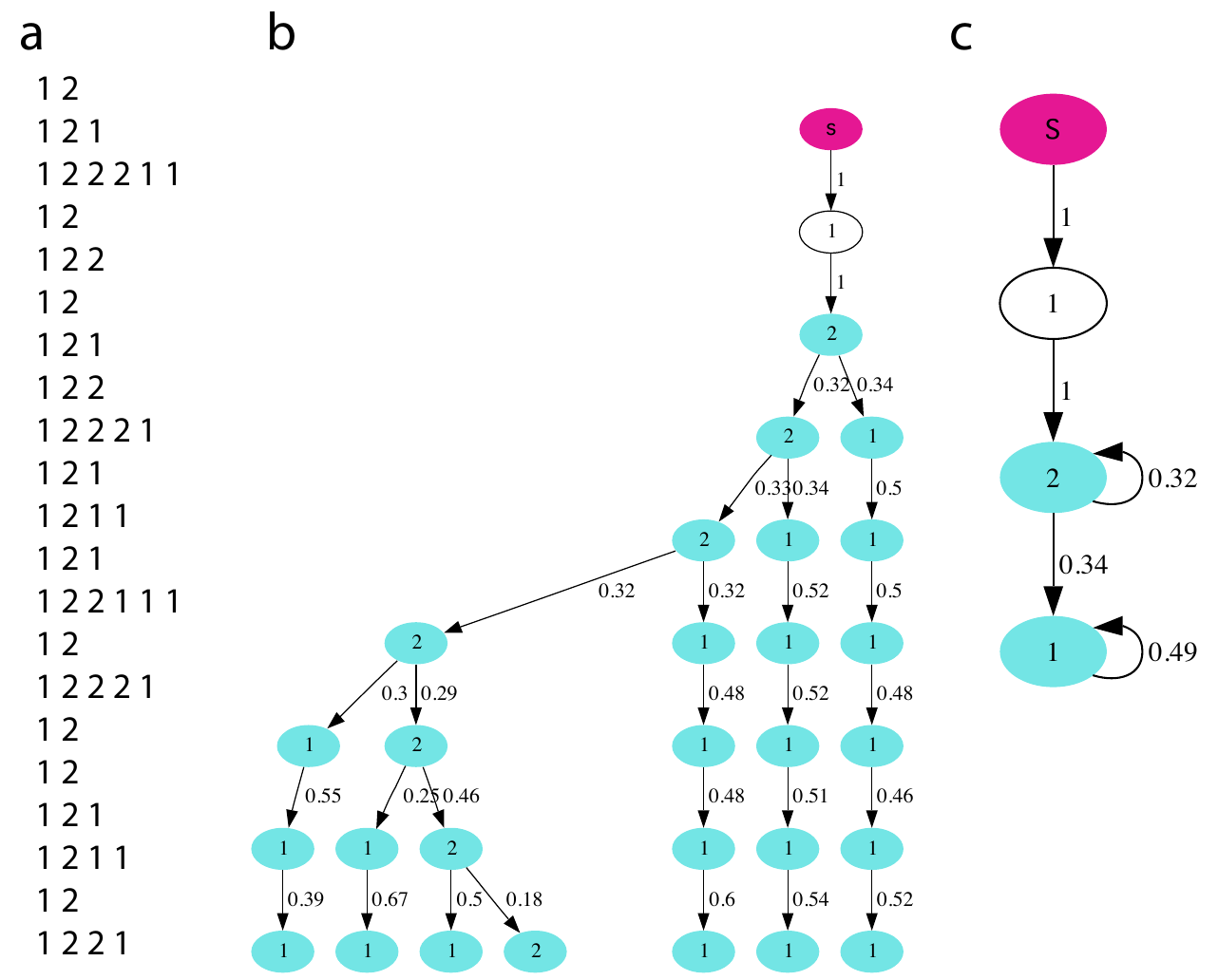}
\end{center}
\caption{ 
{\bf An example of deriving POMM from observed sequences.} 
{\bf a}. The observed sequences generated by a POMM with three states, two states with symbol 1 and one state with symbol 2. 
{\bf b}. The tree-POMM derived from 5000 observed sequences. 
{\bf c}. The derived POMM after merging equivalent states in the tree-POMM. The original model used to generate the sequences shown in {\bf a} are recovered. 
The digram conventions are the same as in Fig.\ref{JinFigSyntaxMarkov}. 
}
\label{JinFigPOMMDerivation}
\end{figure}

To derive a POMM from observed sequences, we developed a state merging method. 
The process is illustrated with an example in Fig.\ref{JinFigPOMMDerivation}
with a simple case of two symbols 1 and 2. 
From 5000 observed sequences (Fig.\ref{JinFigPOMMDerivation}a), 
a tree Markov model is constructed (Fig.\ref{JinFigPOMMDerivation}b). 
For each sequence, the tree model contains a unique path of state transitions from the start state.  
This is achieved by starting with the start state $S_s$ and the end state $S_e$ only, 
and adding new states as needed by finding the paths for the sequences.  
For example, consider the first sequence 12.
At this point no states are emitted from the start state. 
A new state $S_1$ with symbol 1 is added and connected from the start state; a new state $S_2$ with symbol 2 is added and connected from $S_1$; finally, $S_2$ connects to the end state. 
With the additions of the two states, the sequence is mapped to a state transition path $S_s \rightarrow S_1 \rightarrow S_2 \rightarrow S_e$. 
Now consider the second sequence 121. 
State transitions $S_s \rightarrow S_1 \rightarrow S_2$ generate the first two symbols in the sequence. 
To generate the last 1, a new state $S_3$ with symbol 1 is added, 
and is connected from $S_2$ and also to the end state.  
Now $S_2$ branches into $S_3$ and $S_e$. 
This process continues, until all observed sequences are uniqued mapped into the paths in the tree model. 
The transition probabilities from a state to all connected states are computed from the frequencies of the transitions observed in the sequences. 
The tree model is a simple POMM that is a direct translation of the observed sequences; it contains all observed sequences. 
However, the tree model is incapable of generating novel sequences that are statistically consistent with the observed sequences. 
Moreover, since each transition probability can be considered as a parameter, the number of parameters in the tree model is enormous, 
severely restricting its predictive power. 
To reduce the number of parameters, a more concise POMM is derived by merging the equivalent states in the tree model. 
If two states are associated with the same symbol, and the probability distributions of subsequent sequences of length 15 or smaller are similar (cosine-similarity $>0.9$, Materials and Methods), 
the two states are merged. 
This is done until no further mergers are possible. 
Finally, state transitions with probabilities smaller than 0.01 are eliminated, and all states that are reached less than 0.005 times in all observed sequences are also eliminated. 
These merging and pruning procedures lead to a concise POMM with five states for the simple example, as shown in Fig.\ref{JinFigPOMMDerivation}c. 
There are two states for symbol 1, 
which is an example of the many-to-one mapping.  
Indeed, the observed sequences in Fig.\ref{JinFigPOMMDerivation}c
was generated with a POMM with structure identical to the one in Fig.\ref{JinFigPOMMDerivation}c and with equal transition probabilities to all connected states from a given state.  
The example demonstrates that the state merging method can lead to a concise POMM from observed sequences. 

\begin{figure}[!ht]
\begin{center}
\includegraphics[scale=1]{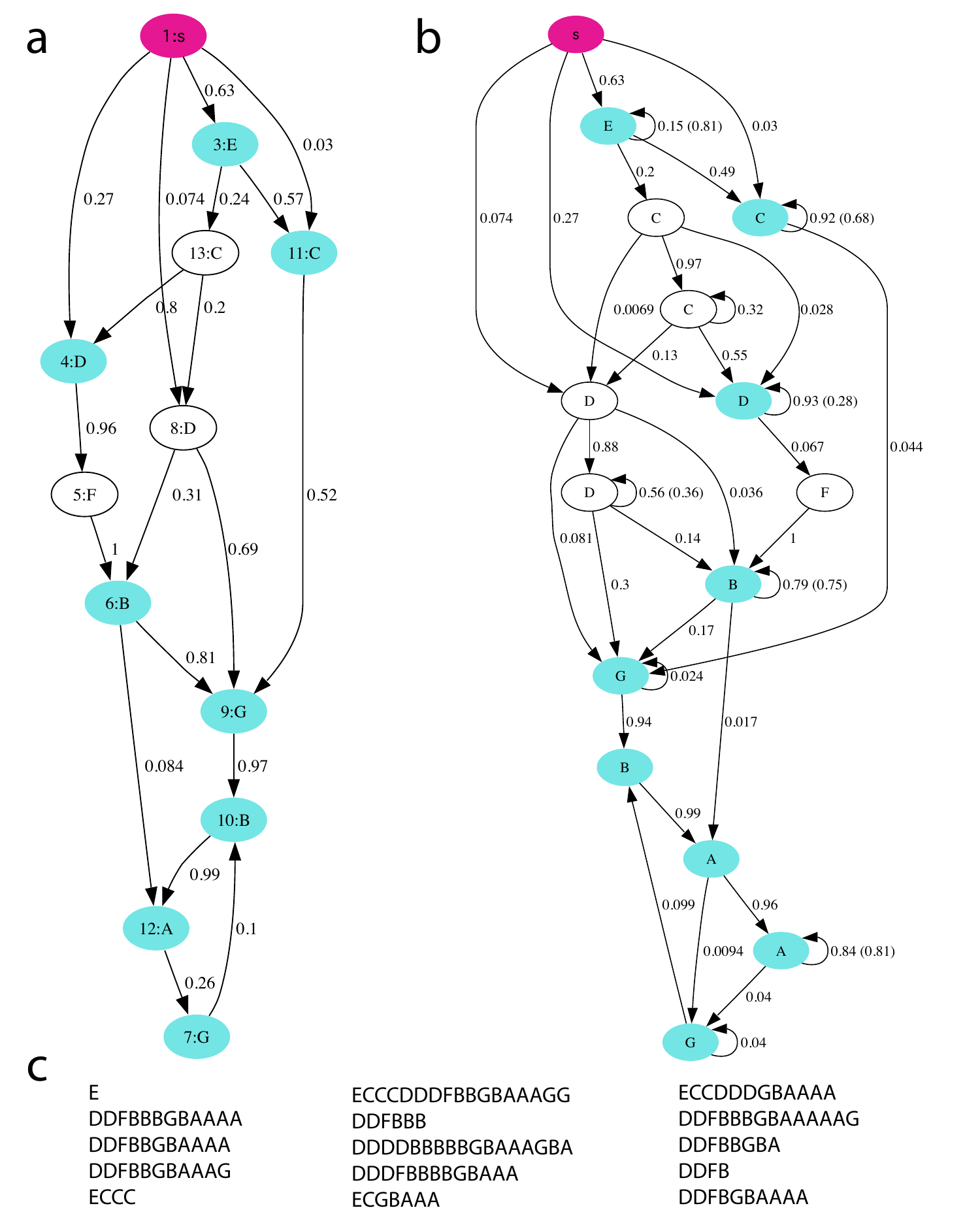}
\end{center}
\caption{ 
{\bf The partially observable Markov model (POMM) syntax.}
{\bf a}. The POMM syntax derived from the observed syllable sequences with syllable repetitions taken out. The numbers in the ovals are state labels. 
{\bf b}. The POMMA syntax derived from the model shown in {\bf a} with the repetitions fitted with adaptation models. 
{\bf c}. Examples of syllable sequences generated from the POMMA shown in {\bf b}. 
The conventions are the same as in Fig.\ref{JinFigSyntaxMarkov}.
}
\label{JinFigSyntaxPOMM}
\end{figure}

We applied the state merging method to the observed Bengalese finch song sequences. 
To incorporate adaptation to syllable repetitions, 
we first derived a POMM with the non-repeat versions of the song sequences, in which the repeats of syllables were ignored but the number of repeats were recorded. 
For example, the non-repeat version of a song sequence ECCDDFBBGBAA is E(1)C(2)D(2)F(1)B(2)G(1)B(1)A(2), 
where the repeat numbers are in the parenthesis. 
While creating the tree model and merging the states, the repeat numbers were kept track of, so that the repeat number distribution for each state could be constructed. 
After following the POMM derivation procedure described above, 
there were 18 states associated with the syllables in the model. 
We further tested deletion of each state and mergers of all pairs of states with the same syllables, while monitoring the fits of the resulting model to the three statistical properties of the non-repeat versions of the song sequences. 
The deletions and mergers were accepted if they did not worsen the fits. 
The resulting POMM, shown in Fig.\ref{JinFigSyntaxPOMM}a, has 11 states associated with the syllables. 
Syllables B, C, D, G are associated with two states each, and syllables A, E, F have one associated state each.  

We next modeled the repeat number distributions in each state with the adaptation model described previously. 
For some states, the adaptation model was not adequate to fit well the repeat number distributions (cosine-similarity of the distributions $<0.95$ with best fitting parameters). 
In such a case, the state $S$ was split into two serially connected states $S_1 \rightarrow S_2$. 
The transitions and associated probabilities to $S$ were set to $S_1$,
and $S_1$ and $S_2$ emitted to the same states and probabilities as $S$. 
$S_2$ has a self-transition with probability $p$ and adaptation parameter $\alpha$, while $S_1$ has no self-transition but has a transition probability $p_1$ to $S_2$.
The repeat number distribution with these parameters is given by 
$P_n = p_1 \alpha^{(n-3)(n-2)/2} p^{n-2} ( 1 - \alpha^{n-2} p)$ (Materials and Methods).  
The parameters were fit with the nonlinear least square fitting procedure.
Each state-splitting thus introduced one more state and one more parameter to the model, and was adequate to fit well the observed repeat number distributions when necessary. 
The resulting POMMA is shown in Fig.\ref{JinFigSyntaxPOMM}b. 
Three states associated with syllables A, C, D were split. 
Altogether, there are 14 states excluding the start and end states, 
and the number of states for syllables A to G are 
2, 2, 3, 3, 1, 1, 2, respectively. 

We generated 10000 syllable sequences from the POMMA (examples shown in Fig.\ref{JinFigSyntaxPOMM}c),  
and compared 
with the observed song sequences the repeat number distributions (Fig.\ref{JinFigRepeatStats}c), the N-gram distributions (Fig.\ref{JinFigNGramStats}c), and the step probabilities (Fig.\ref{JinFigPNStepsStats}c). 
The comparisons are excellent, much better than the Markov model or the Markov model with adaptation. 
Indeed, all $d$-values fall within or close the benchmarks, as shown with the red curves in Fig.\ref{JinFigErrorsSummary}.
Thus, the POMMA captured the statistical properties of the Bengalese finch song. 

\subsection*{Evidence of adaptation} 

\begin{figure}
\begin{center}
\includegraphics[scale=1]{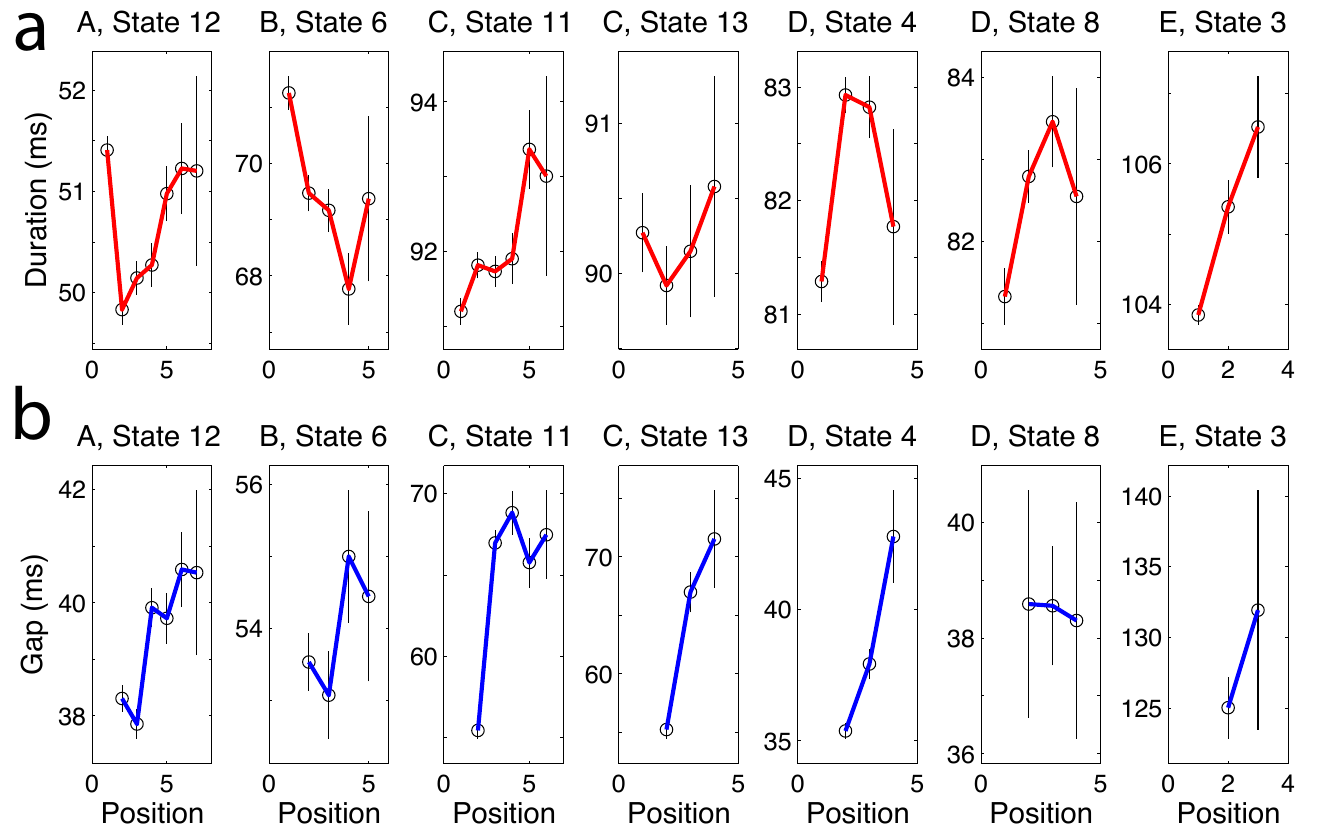}
\end{center}
\caption{
{\bf  Evidence of adaptation.}
The mean durations of syllables ({\bf a}, red lines) and gaps ({\bf b}, blue lines) as functions of the positions in the repeats associated with the states in the POMM
shown in Fig.\ref{JinFigSyntaxPOMM}a. The syllable types and state labels are shown on top of each graph.
The gray lines indicate standard errors. 
}
\label{JinFigAdaptation}
\end{figure}

If the tempos of the syllables are controlled by synaptic chains in HVC \cite{Hahnloser2002,Fee2004,Jin2007,Long2008,Jin2009,Long2010},
adaptations in the neural processes in HVC can lead to lengthening of the syllable durations, just like repetitions of motifs lengthens motif durations in zebra finch \cite{Chi2001}. 
In particular, the durations of repeating syllables should increase with the number of repetitions. 
We tested this idea by examining the durations of repeating syllables in states with repetition numbers more than 3 in the POMM shown in Fig.\ref{JinFigSyntaxPOMM}a. 
The averaged durations of syllables in the same repeat positions, along with the standard errors, are shown in Fig.\ref{JinFigAdaptation}a for syllables A, B, C, D, E. 
There are two states each for syllables C and D. 
The averaged durations and standard errors of the gaps are also plotted in Fig.\ref{JinFigAdaptation}b. 
There are clear indication of duration increase with repetition for syllables C, E, and to some extent D. 
Syllable A has a longer duration in repetition position 1, but then the duration increases steadily after position 2. 
The duration of syllable B decreased with the repetition. 
The gap durations increase with repetition position for all syllables, except that the evidence is weak for syllable D in one state. 
The duration measurements thus provided some evidence for adaptation.  
The more complex behavior of the durations of syllable A might be related to the fact that in the POMMA, two states were needed to accurately describe the repeat number distribution. 
The case of syllable B is a bit puzzling. 
It might be possible that in reality, syllable B is represented with more states than indicated by the POMMA. 

\subsection*{Evidence of many-to-one mapping} 

\begin{figure}[!ht]
\begin{center}
\includegraphics[scale=1]{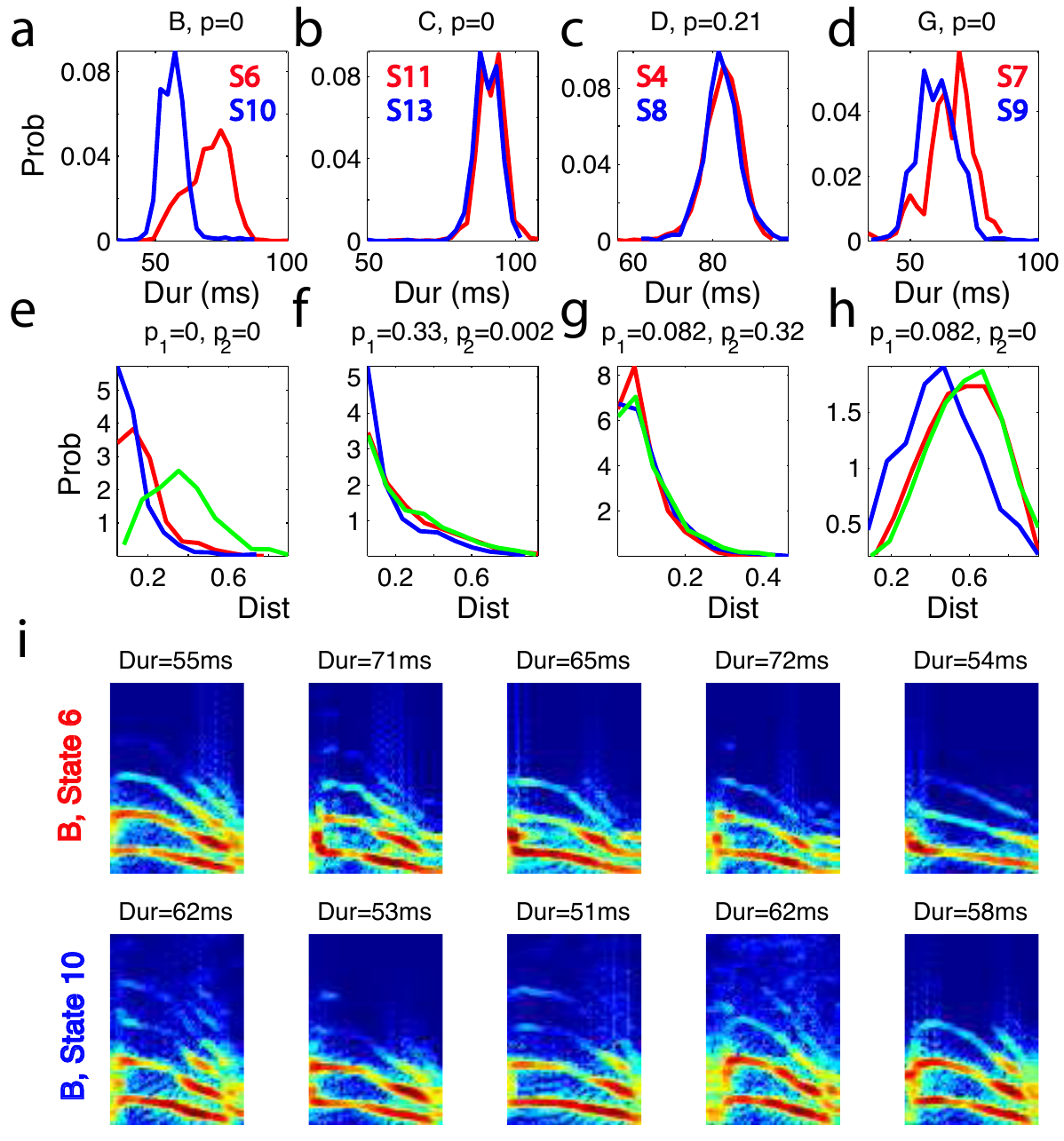}
\end{center}
\caption{
{\bf Evidence of many-to-one mapping from the states to the syllables.}
Syllable durations ({\bf a} to {\bf d}) and 
corresponding distance distributions ({\bf e} to {\bf h}; {\bf e} corresponds to {\bf a}, etc) for the same syllable types associated with different states in the POMM shown in Fig.\ref{JinFigSyntaxPOMM}a. 
In each graph, red and blue curves are from different states. 
The state labels are shown with corresponding colors in {\bf a} to {\bf d}.  
The green curves in {\bf e} to {\bf h} are the distributions of the mutual distances between the syllables in the different states. 
The $p$-values on top of {\bf a} to {\bf d} test the significance that the differences of the means of the two distributions are non-zero. 
The $p_1$ and $p_2$ values on top of {\bf e} to {\bf h} are associated with the red and blue distributions, respectively; 
they are the $p$-values of testing 
the significance that the means of the red or blue distributions have smaller means than the green distributions.  
Graphs {\bf a} to {\bf d} and correspondingly, {\bf e} to {\bf h}, are for syllables B,C,D,G, respectively. 
Panel {\bf i} displays spectrograms of five examples of syllable B associated with state 6 (upper row, corresponds to the red curve in {\bf a}) and five examples associated with state 10 (lower row, corresponds to the blue curve in {\bf a}). 
Durations of the syllables are shown on top. Frequency range of the spectrograms is from $1$-$10kHz$.  
}
\label{JinFigManyToOneMapping}
\end{figure}

In the POMM, different states can be associated with the same syllable type. 
One possible evidence of such many-to-one mapping from states to syllables can be the subtle differences that might exist 
in the syllables associated with different states with the same syllable types.  
There are two states for syllables B,C,D,G in the POMM shown in Fig.\ref{JinFigSyntaxPOMM}a. 
We compared the duration distributions of the same syllable types associated with different states, as shown in Fig.\ref{JinFigManyToOneMapping}a-d. 
The distributions are clearly distinctive for syllables B, C, G ($p = 0$, shuffle test).  
There are no distinctions for syllable D ($p=0.21$). 
We also tested whether syllables in the same states are more similar to each other than to those in different states 
by comparing the distributions of the distances between syllables (Materials and Methods). 
The results are shown in Fig.\ref{JinFigManyToOneMapping}e-h. 
For syllable B, the distance distribution is clearly tighter within the states (red and blue curves in Fig.\ref{JinFigManyToOneMapping}e) than between the states (green curve). 
The mean distances within the states are significantly smaller than the mean distance between the states ($p_1=0, p_2=0$ for the two states, shuffle test). 
For syllable G, the distributions are significantly different for one state ($p_1=0$) but not for the other state ($p=0.08$). 
The same is true for syllable C. 
The distributions are not significantly different for D. 
Despite the significant differences in the durations and distances for syllables B in the two states, the spectrograms of the syllables in the two states are very similar, as shown in Fig.\ref{JinFigManyToOneMapping}i. 
The same is true for other syllables. 
These results provide some evidence for the validity of the many-to-one mapping from the states to the syllables.  

\section*{Discussion}

Bengalese finch song consists of variable sequences of a finite number of syllables. 
We have shown that the statistical properties of the sequences are well captured by a state transition model, POMMA, in which the repeat probabilities of the syllables adapt and many-to-one mappings from the states to the syllables are allowed.  
Similar approach can be useful for characterizing songs of other songbird species with variable syllable sequences. 
The Markov model, which was commonly used in previous studies to characterize the song sequences, is clearly inadequate. 

The POMMA can be directly mapped onto 
the branched chain network model of the Bengalese finch song syntax \cite{Jin2009}.  
Each state of the POMMA corresponds to a syllable-encoding synaptic chain network of HVC projection neurons, and each transition $S_1 \rightarrow S_2$ in the POMMA corresponds to 
the connection from the end of the synaptic chain corresponding to $S_1$
to the start of the synaptic chain corresponding to $S_2$.   
The POMMA and the network model thus have identical branching connection patterns. 
In the network model, 
spike propagation along a chain drives production of a syllable. 
At a branching point, spike propagation continues along one of the connected chain networks with a probability that depends on a winner-take-all competition and noise \cite{Chang2009, Jin2009}. 
Adaptation of repeat probabilities in the POMMA can be related to the reduction of the neuron excitability or the synaptic depression in the chain networks. 
The many-to-one mapping corresponds to the formation of similar connection patterns from different chain networks in HVC to RA neurons such that similar syllables are generated when spikes propagate along these chains. 
The success of the POMMA in capturing the statistical properties of the Bengalese song sequences supports the branched chain network model of Bengalese finch song syntax.
The connection also leads to two critical predictions for the network model. 
One prediction is that HVC networks must have adaptive properties such that repeated activations of a chain network lead to reduction of the self-transition probability. 
It remains to be seen experimentally whether HVC projection neurons or the excitatory synapses between them have such adaptive properties.
It might be also possible to see signatures of adaptation by analyzing the burst intervals of HVC projection neurons during syllable repetitions, or the burst intervals of RA neurons. 
The observation that burst intervals in RA neurons steadily increase with motif repetition in zebra finch \cite{Chi2001} suggests that similar effect could be observed in Bengalese finch.
The other prediction is that, for some syllables, HVC projection neurons should burst intermittently, bursting during some instances of the syllables but not in other instances. 
This is markedly different from the case of zebra finch, in which HVC projection neurons burst reliably for each production of the motif \cite{Hahnloser2002}. 
Our analysis of the syllable durations and distances (Fig.\ref{JinFigAdaptation} and Fig.\ref{JinFigManyToOneMapping}) does provide some evidence for these two predictions. 
But ultimately, they should be tested with electrophysiological experiments. 

We emphasize that 
adaptation is important for reducing the complexity of the state transition model. 
We have attempted to derive a POMM from the observed song sequences without introducing adaptations to the repeat probabilities. 
The number of states, however, expanded to about 40 for the POMM to be able to fit the repeat distributions well enough. 
In particular, many states were needed to produce the peaked repeat number distributions such as that of syllable A (Fig.\ref{JinFigRepeatStats}).  
The significant reduction of the number of states when adaptation is included provides a strong support for its role in shaping the Bengalese finch song syntax. 
We have used multiplicative reduction of the repeat probabilities. 
It remains to be investigated whether other formulations of the adaptation can be similarly or even more effective. 
In our approach, only the repeat probabilities are adapted. 
A more consistent model should allow adaptation and recovery in all transition probabilities, such that the state transition dynamics depends on the history of the entire syllable sequence, not just the syllable repetitions. 
This approach might be important if there are repeats of short sequences such as ABABABAB, in which the transition probabilities from A to B and B to A might need to be adapted.  
But such a model is difficult to derive from the observed sequences. 
In our data, repetitions of short sequences were rarely seen, hence adapting only the repeat probabilities of single syllables is adequate.   

There are previous efforts of describing Bengalese finch song sequences with state transition models \cite{Hosino2000,kakishita2007pattern}. 
Chunks of syllable sequences, which are fixed sequences of syllables, were extracted from the observed sequences and used as the basic units of the state transition models \cite{Hosino2000,kakishita2007pattern}. 
A syllable can appear in many chunks, hence these models implicitly contain the many-to-one mapping from the states to the syllables. 
But the chunk extractions and the state models were not derived from the statistics of the observed sequences. 
Furthermore, the models were not tested against the observed song sequences for statistical properties. 
In contrast, our model is derived from and tested with the observed song sequences. 

There should be a family of equivalent POMMAs for a Bengalese finch song. 
For example, the same repeat distributions can always be modeled with more states. 
The POMMA that we have derived is the simplest model that is consistent with the data.
Given this insight, we expect that the neural representation of the syntax should be similar to the derived POMMA but most likely not identical. 

In conclusion, we have derived a compact POMMA that successfully describes the statistical properties of a Bengalese finch song. 
Our approach can be useful for analyzing other sequential behaviors in animals and statistical properties of sequences in general.

\section*{Materials and Methods}

\subsection*{Vocalization recording}

Acoustic recordings were performed with a boundary microphone (Audio-Technica PRO44). Microphone signals were amplified and filtered (8th-order Bessel high-pass filter with $f_c=300 Hz$ and 8th-order Bessel low-pass filter with $f_c=10 kHz$, Frequency Devices). The filtered signals were digitized with a 16-bit A/D converter (PCI-6251, National Instruments) with a sampling rate of $40 kHz$.

\subsection*{Vocal elements and spectrograms}

From the pressure wave $w(t)$ of a vocalization, the oscillation amplitude $A(t)$ at time $t$ is obtained by finding the maximum of $| w(t) | $  in the interval of one oscillation cycle that contains $t$. 
The amplitude is further transformed to $A_s(t) = S(A(t)^{1/5})$, 
where  $S(\cdot)$ is a smoothing function that uses the second order Savitzky-Golay filter with $20 ms$ window (801 data points). 
Vocal elements are isolated by detecting continuous regions in  $A_s(t)$
that are above a threshold function  $\theta(t)$. 
The threshold function is obtained in $100 ms$ moving windows (step size $5 ms$). 
In each window, the threshold is set at the 0.3 point from the floor $A_{s,min}$  of $A_s(t)$  to the local maximum of $A_s(t)$  in the window. 
The floor $A_{s,min}$  is the characteristic value of $A_s(t)$  in the regimes with no sound, and is identified as the position of the lowest peak in the histogram of the values of $A_s(t)$ for all $t$. 
A detected region is excluded if the total area above $A_{s,min}$ is smaller than $1 ms$ multiplied by the difference between the maximum value $A_{s,max}=\max_t A(t)$  and $A_{s,min}$; 
or if the maximum value of  $A_s(t)$ in the region minus $A_{s,min}$ is smaller than $0.2 ( A_{s,max} - A_{s,min})$; 
or if the width of the region is less than $10 ms$. 
This ensures that most noisy fluctuations are not counted as vocal elements. 

The waveform of an isolated vocal element is transformed into a spectrogram 
$s(f,t)$, which is the energy density at frequency $f$ and time $t$. 
The frequency is restricted to $1 kHz$ to $12 kHz$. 
The spectrogram is computed with the multi-taper method \cite{Mitra2008}
(time-bandwidth product, 1.5; number of tapers, 2) with $5 ms$ window size and $1 ms$ step size (software from http://chronux.org). 
The frequency is discretized into grids with $156.25 Hz$ between adjacent points. To exclude silent periods at the beginning and the end of the vocal element, the time span of the spectrogram is redefined to the region in which the total power in the spectrum at each time point exceeds 5\% of the maximum of the total powers. 

\subsection*{Similarity between vocal elements}

The spectrogram $s(f,t)$ is considered as a sequence of spectra at the discrete time points. 
The spectrum at each time point is smoothed over the frequency domain
using the second order Savitzky-Golay filter with window size of 5 frequency points. 
The smoothed spectrum is further decomposed into 
a slowly varying background $s_b(f,t)$ by smoothing with the second order Savitzky-Golay filter with window size of 20 frequency points; 
and peaks $s_p(f,t)$ by subtracting out $s_b(f,t)$.  
The relative importance of the peaks compared to the background is characterized by the weight
$\alpha_s = s.d.(s_p(f,t))/(s.d.(s_p(f,t)) + s.d.(s_b(f,t)))$, 
where $s.d.$ is the standard deviation of the distribution over the frequency domain. 

The spectrum at $t_1$ of $s_1(f,t_1)$ is compared to 
the spectrum at $t_2$  of $s_2(f,t_2)$ 
by computing 
$$
m_{12} = \alpha C(s_{p,1}(f,t_1),s_{p,2}(f,t_2)) 
+ (1- \alpha) C(s_{b,1}(f,t_1),s_{b,2}(f,t_2)),
$$
which is the weighted sum of the cosine-similarities between the peaks and between the backgrounds. 
Here $s_{p,1}(f,t_1)$ and $s_{p,2}(f,t_2)$ are the peaks
and $s_{b,1}(f,t_1)$ and $s_{b,2}(f,t_2)$
are the backgrounds of $s_1(f,t_1)$ and $s_2(f,t_2)$, respectively. 
The cosine-similarity $C(v_1,v_2)$ of two vectors (or distributions) $v_1,v_2$ are defined as 
$$
C(v_1,v_2) = \frac{(v_1 - \bar{v}_1) \cdot (v_2 - \bar{v}_2)}{
|v_1 - \bar{v}_1| | v_2 - \bar{v}_2| } 
$$
where $\bar{v}_1$ and $\bar{v}_2$ are the means and $| \cdot |$ 
is the norm. 
$\alpha$ is the maximum of the weights across all time points of the two syllables. 
If $m_{12} > 0.75$, the two spectra $s_1(f,t_1)$ and $s_2(f,t_2)$
are considered the same (denoted $s_1(f,t_1) \sim s_2(f,t_2)$). 
Otherwise the two spectra are defined as not the same. 

The similarity between two syllables can be characterized by 
the longest common subsequence (LCS) between them.
A common subsequence is defined by a set of time points
$t_{11} < t_{21} < ... < t_{k1}$ in syllable $s_1(f,t)$
and a set $t_{12} < t_{22} < ... < t_{k2}$ in syllable $s_2(f,t)$, 
such that the spectra at corresponding time points are the same, i.e.
$s_1(f,t_{11}) \sim s_1(f,t_{12})$,  
$s_1(f,t_{21}) \sim s_1(f,t_{22})$,  
...,
$s_1(f,t_{k1}) \sim s_1(f,t_{k2})$.  
There is an additional restriction that corresponding time points do not differ by more than $50 ms$, i.e. $| t_{1j} - t_{2j} | < 50 ms$ for all $1 \le j
\le k $. 
The length of the common subsequence is $k$. 
LCS is the common subsequence with the maximum length. 
A long LCS indicates that the two syllables are similar, while 
a short LCS indicates they are dissimilar. 
We define the similarity score of two syllables as the length of LCS divided by the mean of the lengths of the two syllables. 

\subsection*{Types of vocal elements}

Types of vocal elements are identified by clustering 4000 vocal elements  
using a core-clustering algorithm, modified from the algorithm described in
Jin et al \cite{Jin2009PNAS}. 
The algorithm is based on the distance between vocal elements, 
defined as one minus the similarity score, and consists of the following steps. 
(1) For each vocal element, find the list of nearby vocal elements with distances less than 0.1. 
(2) Among the vocal elements that are not yet part of a cluster, 
select the one with at least 5 nearby vocal elements and the smallest mean distances to its nearby vocal elements 
as the core point of a new cluster. 
(3) All unclustered vocal elements that are in the nearby-list 
of the core point are assigned to the new cluster.
All vocal elements that are in the nearby-list but already clustered are reassigned to the new cluster if their distances to the core points of their respective clusters are larger than their distances to the new core point. 
(4) Repeat steps (2-3) until no new cluster can be created. 
(5) Merge clusters. 
Two clusters are merged if at least 5\% of the vocal elements 
in each cluster have small distances ($<$0.1) to the vocal elements in the other cluster. 
(6) Assign vocal elements that are not yet clustered. 
A vocal element is assigned to the cluster that has the maximum number of members whose distances to the vocal element are less than 0.15.
The spectrograms of all clustered syllables are inspected. 
In some cases, individual clusters can contained separate vocal element types that are have subtle differences but distinguishable. 
Such clusters are split into new clusters. 

Once the types of vocal elements are identified with the clustering algorithm, 
we use the following procedure to classify
all vocal elements that are not already clustered.
(1) Identify the center of each cluster as the vocal element that has the minimum mean distances to all other vocal elements in the cluster. 
(2) Compute the distances between the vocal element to be assigned 
and the cluster centers. 
The three clusters with the lowest distances are selected. 
(3) Compare the durations of the vocal elements in the selected clusters to the duration of the candidate vocal element, and select 20 (or less if the cluster size is smaller than 20) from each selected cluster that are closest.
(4) Compute the distances from the candidate vocal element to the selected vocal elements.
(5) The vocal element is assigned to the cluster to which the most of the selected vocal elements with the distances smaller than 0.2 belong. 
(6) If none of the selected vocal elements have distances less than 0.2, the candidate vocal element is unassigned. 

The unclustered vocal elements are grouped into 2000 blocks, and their mutual distances are computed. 
The clustering and identifying procedures are repeated until no more clusters emerge. 
During this process, clusters are merged if they are subjectively judged as similar by inspecting the spectrograms and the mutual distances between the members of the clusters. 
Individual vocal elements are reassigned to different clusters if necessary. 

\subsection*{Repeats statistics with adaptation}

Consider a state with self-transition. 
The transition probability is $p$ initially but is reduced to $\alpha^{n} p$ after $n$ repetitions of the state, where $0<\alpha <1$ is the adaptation parameter. 
The probability of having $n$ repeats is then 
$$
P_n = p \cdot (\alpha p) \cdot (\alpha^2 p) \cdot ... (\alpha^{n-2} p) 
(1 - \alpha^{n-1} p) 
= \alpha^{(n-2)(n-1)/2} p^{n-1} ( 1 - \alpha^{n-1} p). 
$$

In some cases, it is necessary to use a more complex model with two states to better fit the repeat number distribution. 
We consider a model in which state one transitions to state two with a probability $p_1$, and state two repeats itself with probability $p$ and adaptation parameter $\alpha$. 
The probability of observing one repeat is given by 
$$
P_1 = 1 - p_1.
$$
The probability of observing $n > 1$ repeats is given by 
$$
P_n = p_1 \alpha^{(n-3)(n-2)/2} p^{n-2} ( 1 - \alpha^{n-2} p). 
$$

\section*{Acknowledgments}

We thank Aaron Miller and Jason Wittenbach for reading the manuscript. 

\bibliography{JinKozhevnikovBengalese2010}

\begin{thebibliography}{10}
\providecommand{\url}[1]{\texttt{#1}}
\providecommand{\urlprefix}{URL }
\expandafter\ifx\csname urlstyle\endcsname\relax
  \providecommand{\doi}[1]{doi:\discretionary{}{}{}#1}\else
  \providecommand{\doi}{doi:\discretionary{}{}{}\begingroup
  \urlstyle{rm}\Url}\fi
\providecommand{\bibAnnoteFile}[1]{%
  \IfFileExists{#1}{\begin{quotation}\noindent\textsc{Key:} #1\\
  \textsc{Annotation:}\ \input{#1}\end{quotation}}{}}
\providecommand{\bibAnnote}[2]{%
  \begin{quotation}\noindent\textsc{Key:} #1\\
  \textsc{Annotation:}\ #2\end{quotation}}
\providecommand{\eprint}[2][]{\url{#2}}

\bibitem{Lashley1951}
Lashley KS (1951) The problem of serial order in behavior.
\newblock In: Jeffress LA, editor, Cerebral Mechanisms in Behavior (the Hixon
  Symposium), New York, NY: Wiley. pp. 112-136.
\bibAnnoteFile{Lashley1951}

\bibitem{colonnese1996ontogeny}
Colonnese M, Stallman E, Berridge K (1996) {Ontogeny of action syntax in
  altricial and precocial rodents: grooming sequences of rat and guinea pig
  pups}.
\newblock Behaviour 133: 1165--1195.
\bibAnnoteFile{colonnese1996ontogeny}

\bibitem{Woolley1997}
Woolley SM, Rubel EW (1997) Bengalese finches lonchura striata domestica depend
  upon auditory feedback for the maintenance of adult song.
\newblock J Neurosci 17: 6380-90.
\bibAnnoteFile{Woolley1997}

\bibitem{Honda1999}
Honda E, Okanoya K (1999) Acoustical and syntactical comparisons between songs
  of the white-backed munia (lonchura striata) and its domesticated strain, the
  bengalese finch (lonchura striata var. domestica).
\newblock Zoological Science 16: 319-326.
\bibAnnoteFile{Honda1999}

\bibitem{Okanoya2004}
Okanoya K (2004) The bengalese finch: a window on the behavioral neurobiology
  of birdsong syntax.
\newblock Ann N Y Acad Sci 1016: 724-35.
\bibAnnoteFile{Okanoya2004}

\bibitem{Catchpole1976}
Catchpole C (1976) {Temporal and sequential organisation of song in the sedge
  warbler (Acrocephalus schoenobaenus)}.
\newblock Behaviour 59: 226--246.
\bibAnnoteFile{Catchpole1976}

\bibitem{Todt1998}
Todt D, Hultsch H (1998) {How songbirds deal with large amounts of serial
  information: retrieval rules suggest a hierarchical song memory}.
\newblock Biological Cybernetics 79: 487--500.
\bibAnnoteFile{Todt1998}

\bibitem{Gil2000}
Gil D, Slater P (2000) {Song organisation and singing patterns of the willow
  warbler, Phylloscopus trochilus}.
\newblock Behaviour 137: 759--782.
\bibAnnoteFile{Gil2000}

\bibitem{sakata2006real}
Sakata J, Brainard M (2006) {Real-time contributions of auditory feedback to
  avian vocal motor control}.
\newblock Journal of Neuroscience 26: 9619.
\bibAnnoteFile{sakata2006real}

\bibitem{Jin2009}
Jin D (2009) {Generating variable birdsong syllable sequences with branching
  chain networks in avian premotor nucleus HVC}.
\newblock Physical Review E 80: 51902.
\bibAnnoteFile{Jin2009}

\bibitem{slater1989bird}
Slater P (1989) {Bird song learning: causes and consequences}.
\newblock Ethology Ecology \& Evolution 1: 19--46.
\bibAnnoteFile{slater1989bird}

\bibitem{Hosino2000}
Hosino T, Okanoya K (2000) Lesion of a higher-order song nucleus disrupts
  phrase level complexity in bengalese finches.
\newblock Neuroreport 11: 2091-5.
\bibAnnoteFile{Hosino2000}

\bibitem{kakishita2007pattern}
Kakishita Y, Sasahara K, Nishino T, Takahasi M, Okanoya K (2007) {Pattern
  Extraction Improves Automata-Based Syntax Analysis in Songbirds}.
\newblock In: Progress in artificial life: third Australian conference, ACAL
  2007, Gold Coast, Australia, December 4-6, 2007: proceedings. Springer-Verlag
  New York Inc, p. 320.
\bibAnnoteFile{kakishita2007pattern}

\bibitem{Jurafsky2000}
Jurafsky D, Martin JH (2000) Speech and Language Processing.
\newblock New Jersey: Prentice-Hall.
\bibAnnoteFile{Jurafsky2000}

\bibitem{Chang2009}
Chang W, Jin D (2009) {Spike propagation in driven chain networks with dominant
  global inhibition}.
\newblock Physical Review E 79: 51917.
\bibAnnoteFile{Chang2009}

\bibitem{Sanchez2000}
Sanchez-Vives M, Nowak L, McCormick D (2000) {Cellular mechanisms of
  long-lasting adaptation in visual cortical neurons in vitro}.
\newblock Journal of Neuroscience 20: 4286.
\bibAnnoteFile{Sanchez2000}

\bibitem{Markram1996}
Markram H, Tsodyks M (1996) {Redistribution of synaptic efficacy between
  neocortical pyramidal neurons}.
\newblock Nature 382: 807--810.
\bibAnnoteFile{Markram1996}

\bibitem{Abbott1997}
Abbott L, Varela J, Sen K, Nelson S (1997) {Synaptic depression and cortical
  gain control}.
\newblock Science 275: 221.
\bibAnnoteFile{Abbott1997}

\bibitem{Chi2001}
Chi Z, Margoliash D (2001) Temporal precision and temporal drift in brain and
  behavior of zebra finch song.
\newblock Neuron 32: 899-910.
\bibAnnoteFile{Chi2001}

\bibitem{Hahnloser2002}
Hahnloser RH, Kozhevnikov AA, Fee MS (2002) An ultra-sparse code underlies the
  generation of neural sequences in a songbird.
\newblock Nature 419: 65-70.
\bibAnnoteFile{Hahnloser2002}

\bibitem{Jin2007}
Jin DZ, Ramazanoglu FM, Seung HS (2007) Intrinsic bursting enhances the
  robustness of a neural network model of sequence generation by avian brain
  area hvc.
\newblock J Comput Neurosci 23: 283-99.
\bibAnnoteFile{Jin2007}

\bibitem{Long2008}
Long M, Fee M (2008) {Using temperature to analyse temporal dynamics in the
  songbird motor pathway}.
\newblock Nature 456: 189--194.
\bibAnnoteFile{Long2008}

\bibitem{Long2010}
Long M, Jin D, Fee M (in press) {Support for a synaptic chain model of sequence
  generation from intracellular recordings in the singing bird}.
\newblock Nature .
\bibAnnoteFile{Long2010}

\bibitem{Nottebohm1976}
Nottebohm F, Stokes TM, Leonard CM (1976) Central control of song in the
  canary, serinus canarius.
\newblock J Comp Neurol 165: 457-86.
\bibAnnoteFile{Nottebohm1976}

\bibitem{Nottebohm1982}
Nottebohm F, Kelley DB, Paton JA (1982) Connections of vocal control nuclei in
  the canary telencephalon.
\newblock J Comp Neurol 207: 344-57.
\bibAnnoteFile{Nottebohm1982}

\bibitem{Doya1995}
Doya K, Sejnowski T (1995) {A novel reinforcement model of birdsong
  vocalization learning}.
\newblock Advances in neural information processing systems : 101--108.
\bibAnnoteFile{Doya1995}

\bibitem{Fee2004}
Fee MS, Kozhevnikov AA, Hahnloser RH (2004) Neural mechanisms of vocal sequence
  generation in the songbird.
\newblock Ann N Y Acad Sci 1016: 153-70.
\bibAnnoteFile{Fee2004}

\bibitem{Fiete2004}
Fiete I, Hahnloser R, Fee M, Seung H (2004) {Temporal sparseness of the
  premotor drive is important for rapid learning in a neural network model of
  birdsong}.
\newblock Journal of neurophysiology 92: 2274.
\bibAnnoteFile{Fiete2004}

\bibitem{McCasland1987}
McCasland JS (1987) Neuronal control of bird song production.
\newblock J Neurosci 7: 23-39.
\bibAnnoteFile{McCasland1987}

\bibitem{Yu1996}
Yu AC, Margoliash D (1996) Temporal hierarchical control of singing in birds.
\newblock Science 273: 1871-5.
\bibAnnoteFile{Yu1996}

\bibitem{Callut2004}
Callut J, Dupont P (2004) {A Markovian approach to the induction of regular
  string distributions}.
\newblock Grammatical Inference: Algorithms and Applications : 77--90.
\bibAnnoteFile{Callut2004}

\bibitem{Mitra2008}
Mitra P, Bokil H (2008) {Observed brain dynamics}.
\newblock Oxford University Press, USA.
\bibAnnoteFile{Mitra2008}

\bibitem{Jin2009PNAS}
Jin D, Fujii N, Graybiel A (2009) {Neural representation of time in
  cortico-basal ganglia circuits}.
\newblock Proceedings of the National Academy of Sciences 106: 19156.
\bibAnnoteFile{Jin2009PNAS}

\end{thebibliography}

\section*{Figure Legends}

\end{document}